\documentclass[preprint,aps,prd,superscriptaddress,nofootinbib,showpacs]{revtex4}
\usepackage{graphicx}
\usepackage{amsmath}
\usepackage{amssymb}
\usepackage{multirow}
\usepackage{bm}
\usepackage{color}
\usepackage[hypertex]{hyperref}
\usepackage{relsize}
\RequirePackage{xspace}
\usepackage{dcolumn}

\usepackage{slashed}

\newcommand{\be}{\begin{equation}}
\newcommand{\ee}{\end{equation}}
\newcommand{\ba}{\begin{array}{c}}
\newcommand{\ea}{\end{array}}
\newcommand{\bqa}{\begin{eqnarray}}
\newcommand{\eqa}{\end{eqnarray}}
\newcommand{\bqaa}{\begin{eqnarray*}}
\newcommand{\eqaa}{\end{eqnarray*}}

\begin{document}
\title{\mbox{}\\[10pt]
\textbf{$\chi_{cJ}$ production associated with a $c\bar c$  pair at
hadron colliders}}

\author{Dan Li}
\affiliation{Department of Physics and State Key Laboratory of
Nuclear Physics and Technology, Peking University, Beijing 100871,
China}
\author{Yan-Qing Ma}
\affiliation{Department of Physics and State Key Laboratory of
Nuclear Physics and Technology, Peking University, Beijing 100871,
China}
\author{Kuang-Ta Chao}
\affiliation{Department of Physics and State Key Laboratory of
Nuclear Physics and Technology, Peking University, Beijing 100871,
China}\affiliation{Center for High Energy Physics, Peking
University, Beijing 100871, China}

\date{\today}


\begin{abstract}
$\chi_{cJ}$ (J=0,1,2) production associated with a charm quark pair
in hadron collisions is calculated in the framework of
nonrelativistic QCD at the Tevatron and LHC. It is found that the
color-singlet contribution is small and the color-octet contribution
may be dominant in the large $p_T$ region. The differential cross
section of $p\bar p\to\chi_c+c\bar{c}$ is at least one order of
magnitude smaller than the next-to-leading order result of
$\chi_{cJ}$  inclusive production $p\bar p\to\chi_c+X$, therefore
$\chi_{cJ}$ production in $p\bar p\to\chi_c+c\bar{c}$ may have
negligible influence on the ratio
$R_{\chi_c}=\frac{\sigma_{\chi_{c2}}}{\sigma_{\chi_{c1}}}$ measured
by CDF at the Tevatron. The feeddown contribution from
$\chi_{cJ}+c+\bar{c}$ to $J/\psi+c+\bar{c}$  is found to be large
compared with $J/\psi$ direct production and may have important
influence on the measurement of $J/\psi+c+\bar{c}$. The validity of
fragmentation approximation for the process is also discussed.
\end{abstract}


\maketitle

\section{Introduction}\label{sec:intro}

Charmonium  production associated with a $c\bar{c}$ pair is a good
experimental observable in understanding the production mechanism of
heavy quarkonium. The associated production has been extensive
studied in the literature. In $e^+e^-$ annihilation at B factories,
$J/\psi+c+\bar{c}$ was found to have a very large fraction of
$J/\psi$ inclusive production\cite{Abe:2002PRL,Pakhlov:2002PRD}.
This phenomena can be understood by a large next-to-leading order
(NLO) QCD correction to the color-singlet (CS) $J/\psi+c+\bar{c}$
production\cite{Yu-Jie Zhang:2007PRL}, and a relatively small NLO
QCD correction to the CS $J/\psi+X_{non-c\bar{c}}$
production\cite{Yan-Qing Ma:2009PRL}.  These studies also imply that
the color-octet (CO) contribution to $J/\psi$ production may be very
small and even negligible, and a set of severe constraint on the
linear combination of related CO matrix elements was further
obtained by analyzing $J/\psi$ production in $e^+e^-$
annihilation\cite{Yu-Jie Zhang:2009}. At LEP, in the $Z^0$ decay,
$J/\psi+c+\bar{c}$\cite{Rong Li:2010} was found to be the dominant
contribution to the $J/\psi$ inclusive
production\cite{Abreu:1994PLB}. On the contrary, in  $\gamma\gamma$
collisions, the contribution of $J/\psi$ associated production
\cite{Rong Li:2009PRD} was estimated to be several orders of
magnitude smaller than the experimental data of $J/\psi$ inclusive
production\cite{Abdallah:2003PLB}. In hadron collisions at the
Tevatron, theoretical
predictions\cite{Artoisenet:2007PLB,zhiguoHe:2009PRD} showed that
the $J/\psi+c+\bar{c}$ contribution was significant in the large
$p_T$ region compared with the NLO result of non-$c\bar{c}$
contributions, and the produced $J/\psi$ is mainly unpolarized,
which is analogous to the polarization of $J/\psi$ in inclusive
production\cite{Campbell:2007,Abe: 1997PRL}. The integrated cross
section was also significant and showed a great measurable potential
both at the Tevatron and RHIC. $J/\psi$ associated production was
also considered in the $\Upsilon$ decay\cite{Zhi-Guo He:2009} to
explore the CO mechanism in heavy quarkonium decays. As shown, most
of those studies focused on $J/\psi$ associated production. However,
due to the importance of charmonium associated production, studies
of associated production for other charmonium states other than
$J/\psi$ may also be valuable.

In this paper, we perform a calculation for the P-wave charmonium
$\chi_{cJ}$ (J=0,1,2) associated production in hadron collisions in
the framework of nonrelativistic QCD(NRQCD)\cite{Bodwin:1994jh}. The
motivation for this work is two fold. The first is related to the
ratio $R_{\chi_c}=\frac{\sigma_{\chi_{c2}}}{\sigma_{\chi_{c1}}}$
measured by the CDF collaboration at the Tevatron. CDF found that
$R_{\chi_c}$  approaches to about 0.75 at large
$p_T$\cite{Abulencia:2007PRL}. However, if the $\chi_{cJ}$ inclusive
production is dominated by the CO process (because the CS
contribution at large $p_T$ is suppressed by $\frac{1}{p_T^6}$ at
leading order(LO)), the value of $R_{\chi_c}$ should tend to be
$\frac{5}{3}$, which is predicted by naive spin counting. Recently,
the calculation of NLO QCD correction to $\chi_{cJ} + X_{non-c\bar
c}$ inclusive production is performed\cite{Yan-Qing Ma:2010} and it
is found that the NLO correction for CS channel can bring out a
$\frac{1}{p_T^4}$ term, which makes the CS contribution much
important at large $p_T$, and then by combining CS with CO
contributions one is able to fit the experimental value of
$R_{\chi_c}$  quite well over a wide $p_T$ region. The $\chi_{cJ}$
associated production with a $c\bar c$ pair is of the same order in
pQCD as the NLO $\chi_{cJ}+X$ inclusive production and it contains
also fragmentation contributions which scale as $\frac{1}{p_T^4}$.
So it is interesting to check whether the $\chi_{cJ}c\bar c$
associated production is also very large and whether it can further
improve theoretical predictions of ratio $R_{\chi_c}$.   The second
reason is that the measurements at the Tevatron for the production
rates of $J/\psi+c+\bar{c}$ and $J/\psi+X$ are important on shedding
light on understanding the $J/\psi$ production mechanism in hadron
collisions. And the prompt $J/\psi$ production receives significant
feeddown contributions from $\psi(2S)$ and $\chi_{cJ}$. So it is
important to know how large is the feeddown contribution to $J/\psi$
associated production from $\chi_{cJ}$ associated production. The
result itself in this work can also give information for directly
measuring $\chi_{cJ}$ from $\chi_{cJ}+c+\bar{c}$ production at
hadron colliders.

The remainder of this paper is organized as follows. In Sec.
\ref{sec:calc} we briefly describe our calculation method. In Sec.
\ref{sec:resu} we give our numerical results and analyze the
obtained results. In the last section, we give a summary.

\section{Calculation of $p \bar{p}\longrightarrow \chi_{cJ}+c+\bar{c}$}\label{sec:calc}

\begin{figure}[htb]
\begin{center}
\scalebox{0.7}{\includegraphics{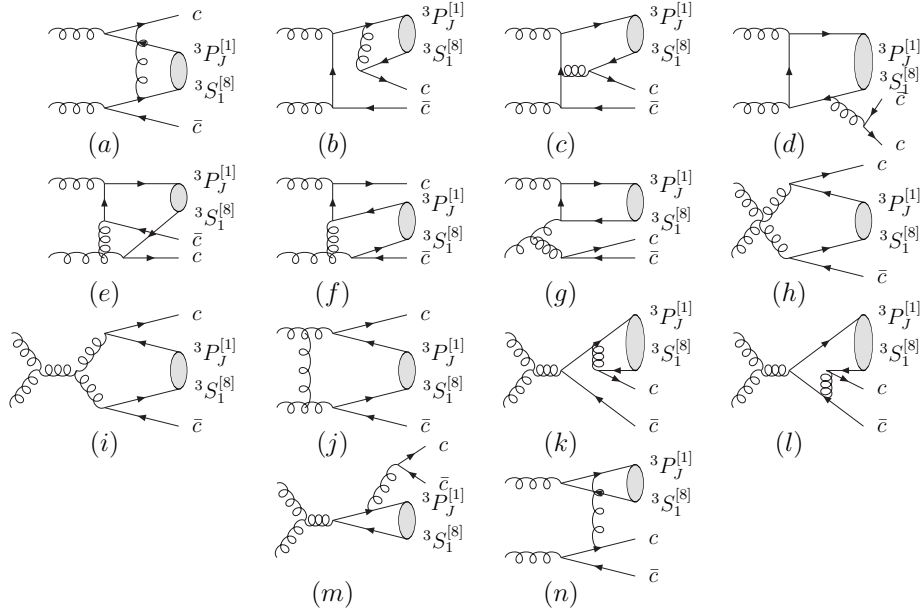}}
\end{center}
\caption{\label{fig:1}Representative Feynman diagrams for $p\bar p
\to\chi_{cJ}+c+\bar{c}$.}
\end{figure}

\begin{figure}[htb]
\begin{center}
\scalebox{0.7}{\includegraphics{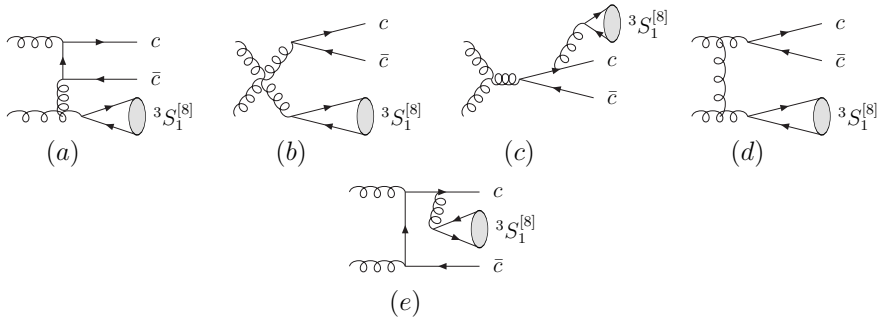}}
\end{center}
\caption{\label{fig:2} Representative extra diagrams for CO
channel.}
\end{figure}

In the framework of NRQCD factorization, the cross section for the
$\chi_{cJ}c\bar c$ associated production in proton-antiproton
collisions has the following form:
\begin{align}
{\rm d}\sigma[p\bar{p}\rightarrow \chi_{cJ}+c+\bar{c}]=&\sum_{i,j,n}
\int {\rm d}x_1 {\rm d}x_2 f_{i/p}(x_1)f_{j/\bar{p}}(x_2)\nonumber\\
&\times {\rm d}\hat{\sigma}(i+j\rightarrow
c\bar{c}[n]+c+\bar{c})\langle O^{\chi_{cJ}}[n]\rangle
\end{align}
where i, j denote the initial state partons from the proton or
anti-proton. We assume the contribution from light quark
annihilation to be negligible,
so i, j are gluons in our case. The quantum numbers $n$ represent
the color and orbital angular momentum of the intermediate $c\bar c$
states at short-distances, which evolve into the $\chi_{cJ}$ meson
at long-distances. At leading order in relative velocity $v$ of the
$c\bar{c}$ pair, $n$ can be taken as $n={}^3P_J^{[1]}$ for the CS
and $n={}^3S_1^{[8]}$ for the CO intermediate states. We use
\textbf{FeynArts}\cite{Hahn:2001} to generate Feynman diagrams. For
the CS case, there are 56 Feynman diagrams, of which some
representative are shown in Fig. \ref{fig:1}. For the CO case, there
are 16 extra Feynman diagrams relative to the CS case and their
extra topology structures are shown in Fig. \ref{fig:2}. These extra
topology structures actually represent gluon fragmentation
contributions.

The fragmentation diagrams can give a $\frac{1}{p_T^4}$ contribution
for the differential cross section $\frac{{\rm d}\sigma}{{\rm
d}y{\rm d}p_T^2}$, while remained diagrams can at most give
$\frac{1}{p_T^6}$ terms. Therefore, one may expect the fragmentation
contribution is dominant at large $p_T$. Note that, however, not
only the CO channel ($^3S_1^{[8]}$) have fragmentation contribution,
but the CS channel has also quark (anti-quark) fragmentation
contributions. In Fig.\ref{fig:1}, the diagrams (b) and (k)
represent (anti-)quark fragmentation contributions in the large
$p_T$ approximation, where one of the charm or anti-charm quark
fragments to $\chi_{c}$ plus another charm or anti-charm quark.
However, only when $p_T$ is large enough, these quark diagrams can
reach the fragmentation region and give leading contributions.
Otherwise, these quark diagrams can give only small contributions.
We will discuss the fragmentation approximation in the next section.

We use the spinor-helicity method to deal with Feynman
amplitudes\cite{Chang:2004cpc} and use the package
\textbf{S@M}\cite{Maitre:2007} to simplify the Feynman amplitudes in
spinor form. Based on this method, the spinor form for massive
external fermions can be written as
\begin{align}
u_{\pm\frac{1}{2}}(p)&=\frac{1}{\sqrt{2p\cdot q}}(\slashed{p}+m)|q_{0\pm}\rangle, \\
v_{\pm\frac{1}{2}}(p)&=\frac{1}{\sqrt{2p\cdot
q}}(\slashed{p}-m)|q_{0\mp}\rangle,
\end{align}
where $u_{\pm\frac{1}{2}}(p)$($v_{\pm\frac{1}{2}}(p)$) are Dirac
spinors of massive (anti-)fermion with momentum $p$ and spin
$\pm\frac{1}{2}$; $q_{0\pm}$ are reference Weyl spinors with
light-like reference momentum $q_{0}$ and helicities $\lambda=\pm1$.
In principle, $\overrightarrow{\textbf{q}}_0$ should be along the
axis of $\overrightarrow{\textbf{p}}$ to guarantee the validity of
above equations for individual spin. However, if we just concern
about a result by summing over spin,  $q_0$ can be chosen
arbitrarily. Here, it is chosen to be one of the initial partons'
momentum in order to simplify the calculation. The polarization
vectors for external gluon with momentum $k$ and light-like
reference momentum $q_0$ are represented as
\begin{align}
\slashed{\epsilon}^+(k,q_0)=&\frac{\sqrt{2}}{\langle q_0| k\rangle
}(|k_-\rangle \langle q_{0-}|+|q_{0+}\rangle \langle
k_+|),\\
\slashed{\epsilon}^-(k,q_0)=&\frac{\sqrt{2}}{\langle q_0| k\rangle
^*}(|k_+\rangle \langle q_{0+}|+|q_{0-}\rangle \langle k_-|).
\end{align}
Using the following identity, we decouple the spin projection
operator\cite{Dan:2009PRD} for bound states from Feynman amplitudes:
\begin{eqnarray}
\sum_{\lambda_{2}\lambda_{3}}(\slashed{p}_{\bar{c}}-m_c)|q_{0\lambda_{2}}\rangle
\langle q_{0\lambda_{2}}|P_{1S_z}|q_{0\lambda_{3}}\rangle \langle
q_{0\lambda_{3}}|(\slashed{p}_c-m_c)=2p_c\cdot q_0 2p_{\bar{c}}\cdot
q_0 P_{1S_z},
\end{eqnarray}
where there is a relative velocity $v$ between $p_c$ and
$p_{\bar{c}}$ for the P-wave case. Then with the help of Fierz
transformation (and its generalized forms), the amplitudes are
reduced to
\begin{eqnarray}
M_i=C_i^{jk} f_j f_k,
\end{eqnarray}
where $i$ is the index for different diagrams and $f_j$ are the
simplified fermion chains, and the three-gluon vertex are properly
dealt with (see \cite{Chang:2004cpc} for details). The specific
expressions of $f_j$ encountered here are listed in the appendix.

We write the polarization tensors for $\chi_{cJ}$ explicitly. For
$\chi_{c0}$ it is symmetric for the two indexes $\mu$$\nu$:
\begin{eqnarray}
\epsilon^{\mu\nu}=-g_{\mu\nu}+\frac{p_\mu p_\nu}{m_c^2};
\end{eqnarray}for $\chi_{c1}$ it is
anti-symmetric for the two indexes $\mu$$\nu$:\begin{align}
\epsilon^{\mu\nu}_1=&\frac{1}{\sqrt{2}}(\epsilon^\mu_x
\epsilon^\nu_y-\epsilon^\nu_x
\epsilon^\mu_y),\\
\epsilon^{\mu\nu}_2=&\frac{1}{\sqrt{2}}(\epsilon^\mu_x
\epsilon^\nu_z-\epsilon^\nu_x
\epsilon^\mu_z),\\
\epsilon^{\mu\nu}_3=&\frac{1}{\sqrt{2}}(\epsilon^\mu_y
\epsilon^\nu_z-\epsilon^\nu_y \epsilon^\mu_z);
\end{align}for $\chi_{c2}$ it is again
symmetric for the two indexes $\mu$$\nu$:\begin{align}
\epsilon^{\mu\nu}_1=&\frac{1}{\sqrt{2}}(\epsilon^\mu_x
\epsilon^\nu_y+\epsilon^\nu_x
\epsilon^\mu_y),\\
\epsilon^{\mu\nu}_2=&\frac{1}{\sqrt{2}}(\epsilon^\mu_x
\epsilon^\nu_z+\epsilon^\nu_x
\epsilon^\mu_z),\\
\epsilon^{\mu\nu}_3=&\frac{1}{\sqrt{2}}(\epsilon^\mu_y
\epsilon^\nu_z+\epsilon^\nu_y \epsilon^\mu_z),\\
\epsilon^{\mu\nu}_4=&\frac{1}{\sqrt{2}}(\epsilon^\mu_x
\epsilon^\nu_x-\epsilon^\nu_y
\epsilon^\mu_y),\\
\epsilon^{\mu\nu}_5=&\frac{1}{\sqrt{6}}(\epsilon^\mu_x
\epsilon^\nu_x+\epsilon^\mu_y \epsilon^\nu_y-2\epsilon^\mu_z
\epsilon^\nu_z).
\end{align}
The definition of $\epsilon_x$, $\epsilon_y$, and $\epsilon_z$
are\begin{align}
\epsilon_x(P)=&(0,\cos \theta\cos \phi,\cos \theta \sin \phi ,-\sin \theta ),\\
\epsilon_y(P)=&(0,-\sin \phi ,\cos \phi ,0),\\
\epsilon_z(P)=&\frac{1}{M}(|\overrightarrow{P}|,P_0\sin \theta \cos
\phi ,P_0\sin \theta \sin \phi ,P_0\cos \theta),
\end{align}
where M, $\overrightarrow{P}$, and $P^0$ are the mass, momentum and
energy of $\chi_{cJ}$; angles $\theta$ and $\phi$ describe
$\chi_{cJ}$'s direction\cite{Chao-Hsi Chang:2004PRD}. For CO
${}^3S_1^{[8]}$, the spinor-helicity forms of polarization vectors
are kept as $\langle
q_{0\lambda}|\slashed{\varepsilon}|q_{0\lambda}\rangle $, $\langle
q_{0\lambda}|\slashed{\varepsilon}\slashed{p}|q_{0\lambda}\rangle $,
$\langle
q_{0\lambda}|\slashed{p}\slashed{\varepsilon}|q_{0\lambda}\rangle $
until numerically squaring the amplitudes in Fortran program.

In our numerical calculation, we give only a rapidity cut condition
for $\chi_c$. However, to detect the associated production, one
should detect at least another hadron containing charm or anti-charm
quark, and the rapidity cuts from experimental facility should also
apply for the (anti-)charm quarks in principle.

For phase space integration, we use the general $2\rightarrow 3$
phase space expression, plus two fold momentum fraction integration
for initial partons:
\begin{eqnarray}
\frac{{\rm d}\sigma}{{\rm d}p_T}=\int_{\delta}^1 {\rm
d}x_1f_{g/p}(x_1)\int_{\delta/x_1}^1 {\rm
d}x_2f_{g/\bar{p}}(x_2)\int_{\sqrt{m_5^2+p_T^2}}^{(k_5^0)_{max}}{\rm
d}k_5^0\int_{(k_3^0)_{min}}^{(k_3^0)_{max}}{\rm d}k_3^0
\int_0^{2\pi}{\rm d}\eta \sum|M|^2,
\end{eqnarray}
where $\delta=\frac{16m^2+4p_T^2}{s}$, $k_5^0$ is the energy of
$\chi_c$, $k_3^0$ is the energy of one of the emitted charm or
anti-charm quark, $\eta$ describes the angle between the plane for
the final three particles and the plane chosen for the two initial
partons, and we omit flux factor and other normalization factors.
The upper and low limits for $k_5^0$ and $k_3^0$ integration are a
little complicated so we don't list them here. We use Vegas
\cite{Lepage:Vegas} in Fortran program to perform the numerical
integration. The correctness of our phase space integration program
is verified by comparing the calculated $J/\psi+c+\bar{c}$
production (we calculate it again) with the result from
Ref.\cite{Artoisenet:2007PLB}.

\section{Result and analysis}\label{sec:resu}

In numerical calculation, we choose $m_c=1.5\text{GeV}$. The
factorization scale and renormalization scale are both chosen as
$\mu_0=\sqrt{p_T^2+4m_c^2}$. We use CTEQ6M as PDF input. The CS
matrix element $\langle O^{\chi_{cJ}}[{}^3P_J^{[1]}]\rangle $ is
related to the P-wave function at the origin by the formula:
$\langle O^{\chi_{cJ}}[{}^3P_J^{[1]}]\rangle
=\frac{(2J+1)3N_c}{2\pi}|R_P^{\prime}(0)|^2$ and we
choose:$|R_P^{\prime}(0)|^2=0.075\text{GeV}^{5}$ from the potential
model calculations\cite{Eichten:1995PRD}. For the CO matrix element
$\langle O^{\chi_{cJ}}[{}^3S_1^{[8]}]\rangle $, by spin symmetry, we
have the following relation that $\langle
O^{\chi_{c0}}[{}^3S_1^{[8]}]\rangle :\langle
O^{\chi_{c1}}[{}^3S_1^{[8]}]\rangle :\langle
O^{\chi_{c2}}[{}^3S_1^{[8]}]\rangle = 1:3:5$ and we use $\langle
O^{\chi_{c0}}[{}^3S_1^{[8]}]\rangle \approx 2.2\times10^{-3}
\text{GeV}^3$ as the central value obtained from fitting the
$\chi_{cJ}$ inclusive production $p\bar{p}$ to
$\chi_{cJ}+X_{non-c\bar{c}}$ at the Tevatron\cite{Yan-Qing Ma:2010}.

\begin{figure}[!]
\begin{center}
\scalebox{0.55}{\includegraphics{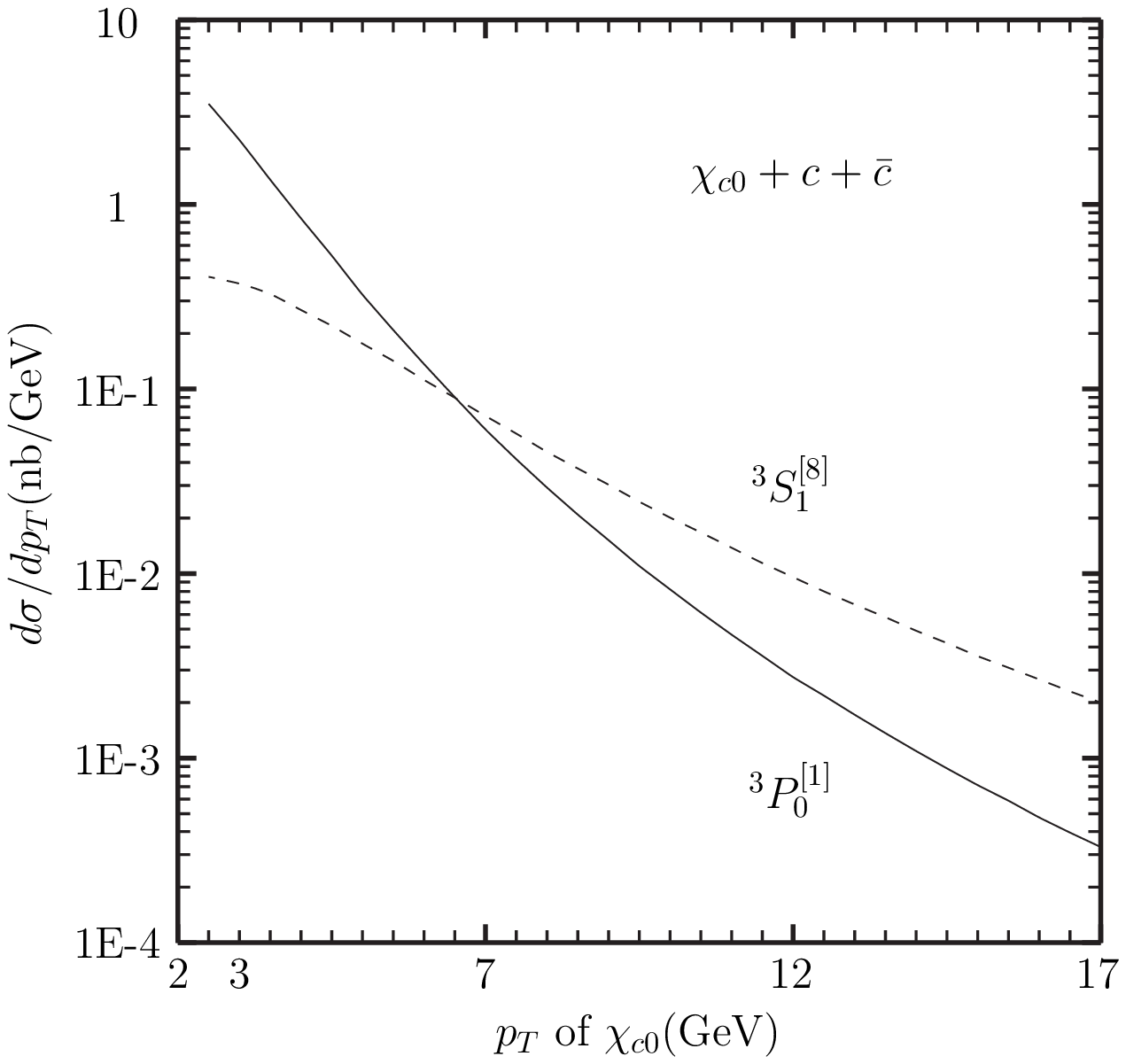}}\\
\scalebox{0.55}{\includegraphics{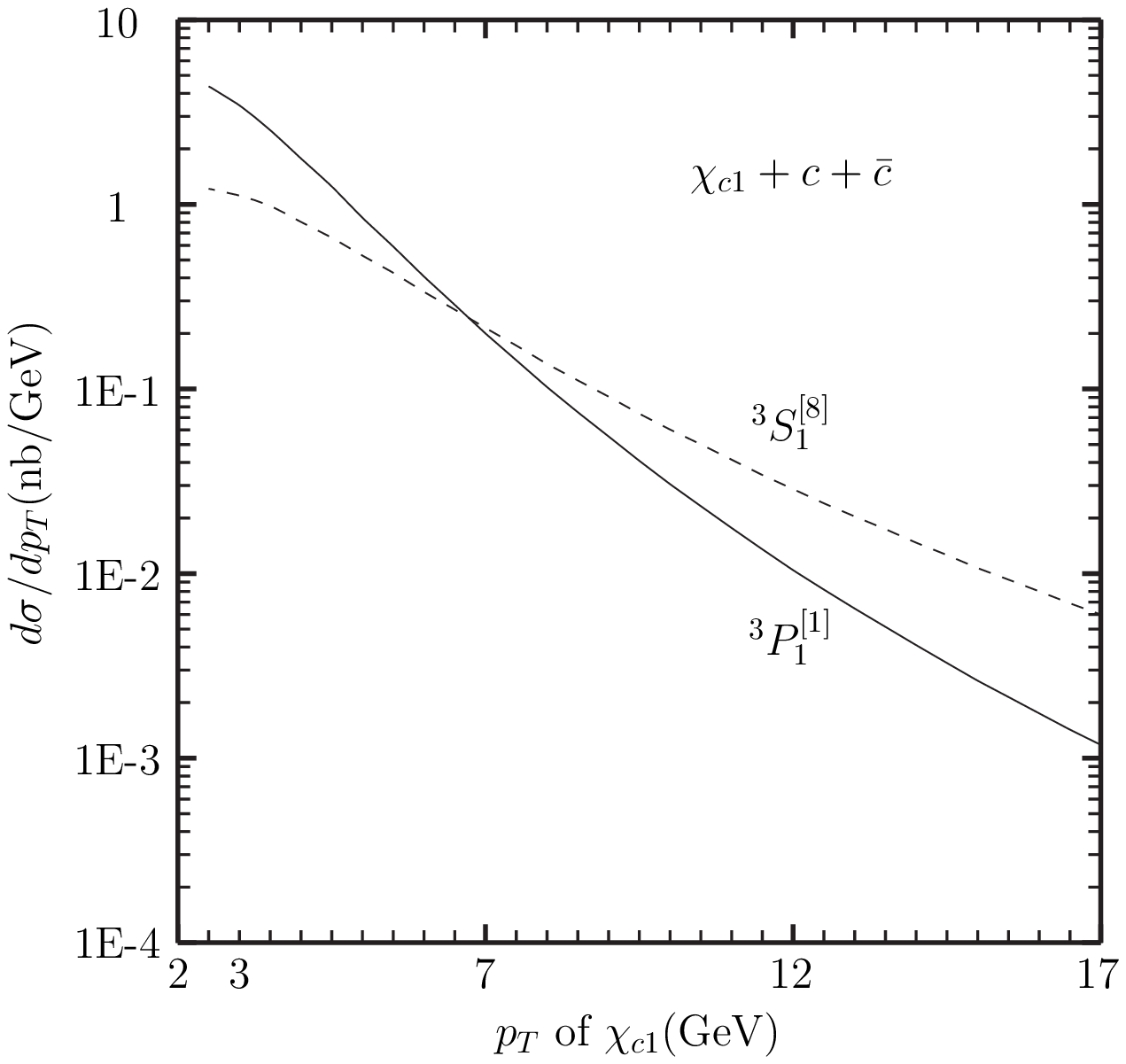}}\\
\scalebox{0.55}{\includegraphics{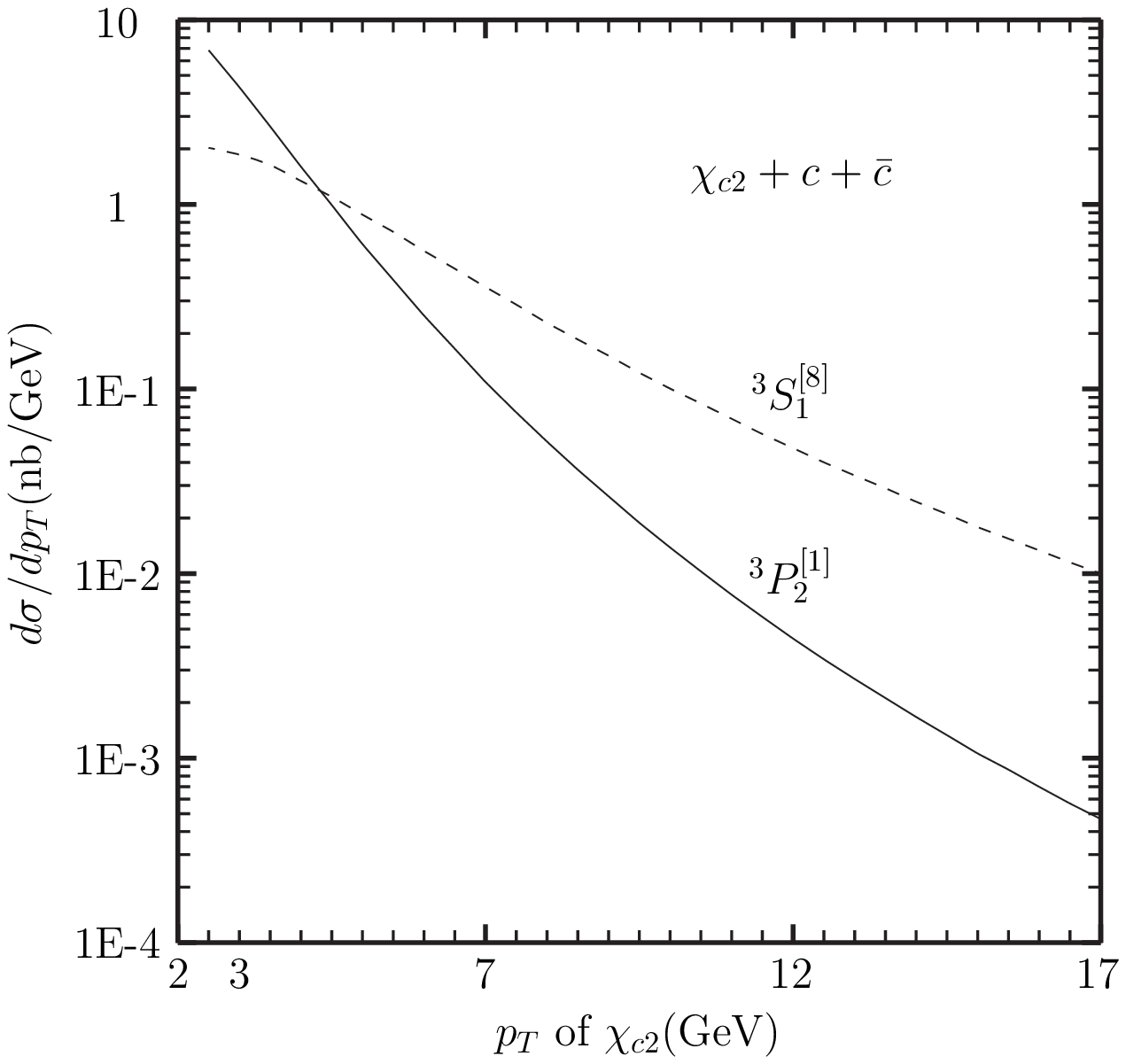}}
\end{center}
\caption{\label{fig:3}Differential cross sections of
$\chi_{cJ}+c+\bar{c}$ as functions of transverse momentum at the
Tevatron with $\sqrt{s}=1.96\text{GeV}$ and rapidity cut
$|y_{\chi_{cJ}}|<0.6$. The dashed line denotes CO contribution, and
the solid line denotes CS contribution.}
\end{figure}

In Fig. \ref{fig:3},
we show both CS and CO contributions to the differential cross
section of $\chi_c+c+\bar{c}$. It is found that the CO contribution
dominates over production in the large $p_T$ region, and it
decreases much slower than that of CS as $p_T$ increases. This seems
to conflict with that both of the two channels should behave as
$\frac{1}{p_T^4}$ at large $p_T$. To see this point more clearly, we
fit the parton differential cross section (taking out the influence
of PDF) with $\frac{1}{p_T^n}$, and it turns out that the CS channel
scales roughly as $\frac{1}{p_T^6}$ while the CO channel scales as
$\frac{1}{p_T^4}$ in the region $p_T<17$ \text{GeV}. This implies
that the CO contribution in Fig. \ref{fig:2} is dominated by gluon
fragmentation just as expected, but the CS channel has not reached
the (anti-)quark fragmentation region, and its contribution is still
suppressed in the moderately large $p_T$ region (e.g., $p_T\leq 17$
\text{GeV}).

In order to understand the above mentioned $p_T$ behavior, we study
the $J/\psi+c\bar{c}$ production as an example. We calculate
(anti-)quark fragmentation diagrams in axial gauge to include all
$\frac{1}{p_T^4}$ contributions. $\sqrt{s}$ can be set to be 100TeV
to enable us to calculate at $p_T$ as large as possible. We find
that when $p_T= 50\text{GeV}$, (anti-)quark fragmentation
contribution has a fraction of about 70\% of the total differential
cross section, and then the fraction rises to about 93\% when
$p_T=150\text{GeV}$. And the fraction reaches 100\% (within the
calculation errors) when $p_T>450\text{GeV}$. Thus we find that
(anti-)quark fragmentation approximation is only valid for very
large $p_T$ ($p_T>100\text{GeV}$ at least), while for the presently
interested $p_T$ region ($p_T \lesssim 17\text{GeV}$) the
$\frac{1}{p_T^4}$ term induced by (anti-)quark fragmentation is very
small and not important. This explains the fact that the CS channels
in $\chi_{cJ}$ production behave almost as $\frac{1}{p_T^6}$.

\begin{figure}[!]
\begin{center}
\scalebox{0.55}{\includegraphics{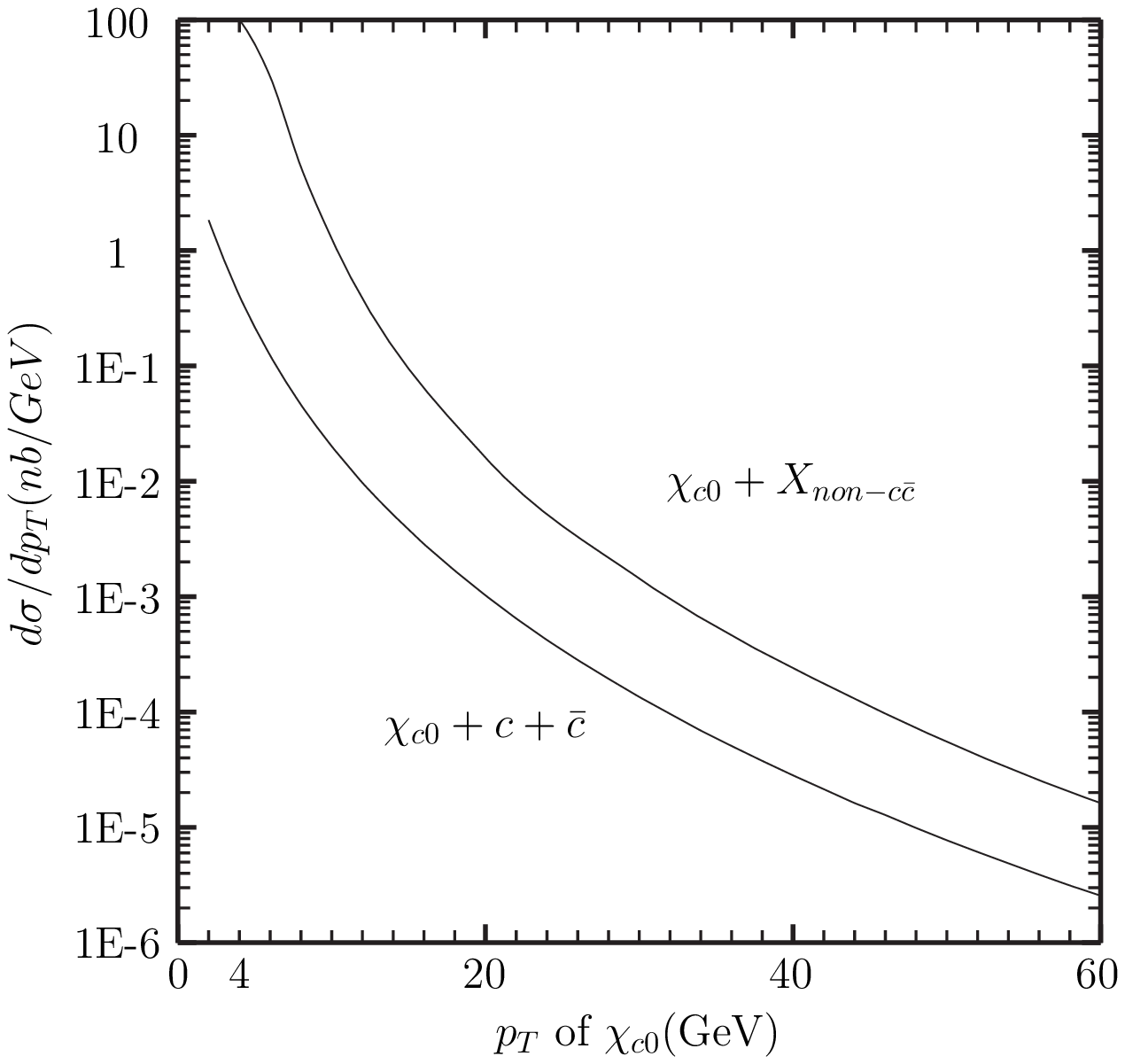}}\\
\scalebox{0.55}{\includegraphics{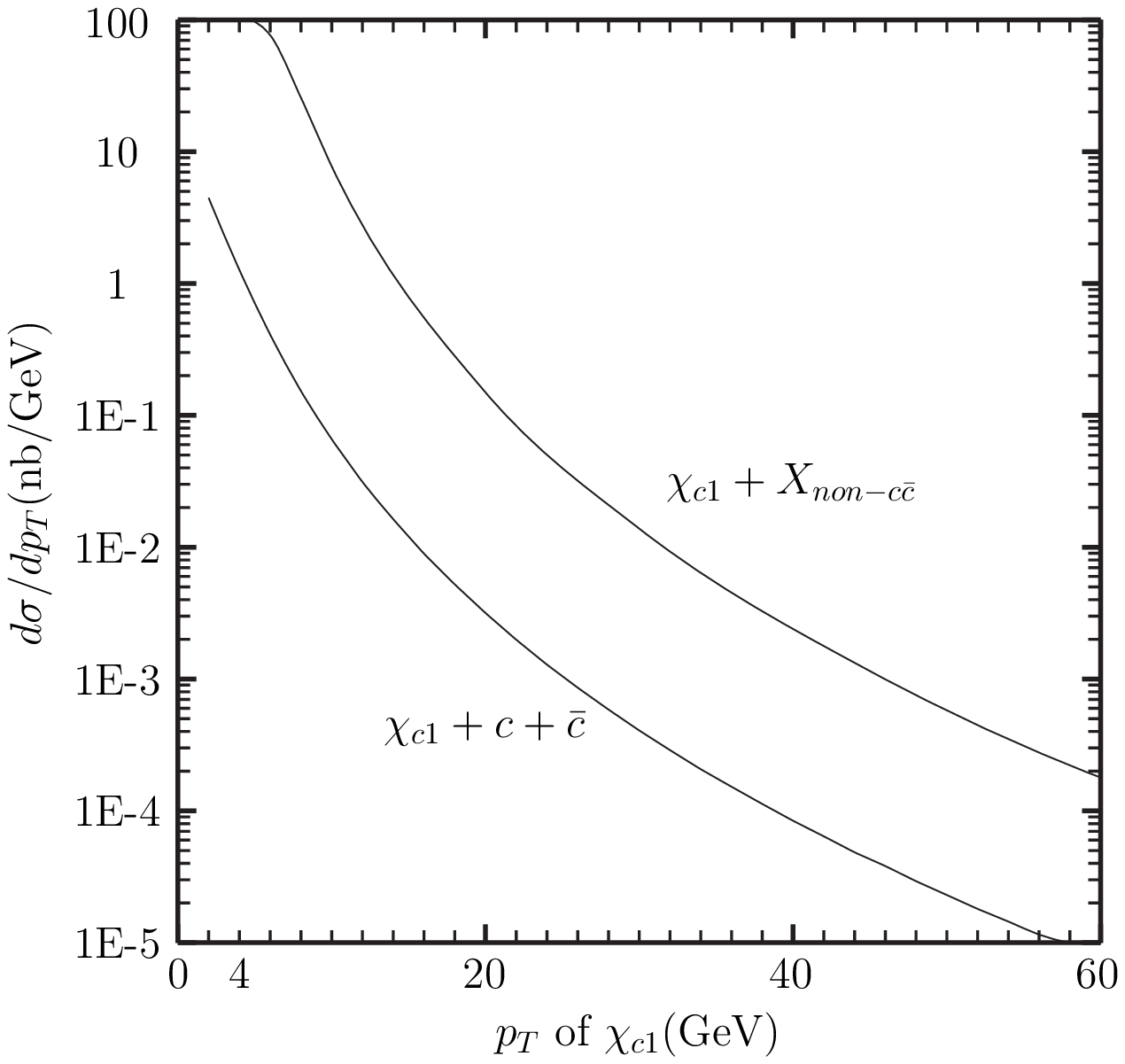}}\\
\scalebox{0.55}{\includegraphics{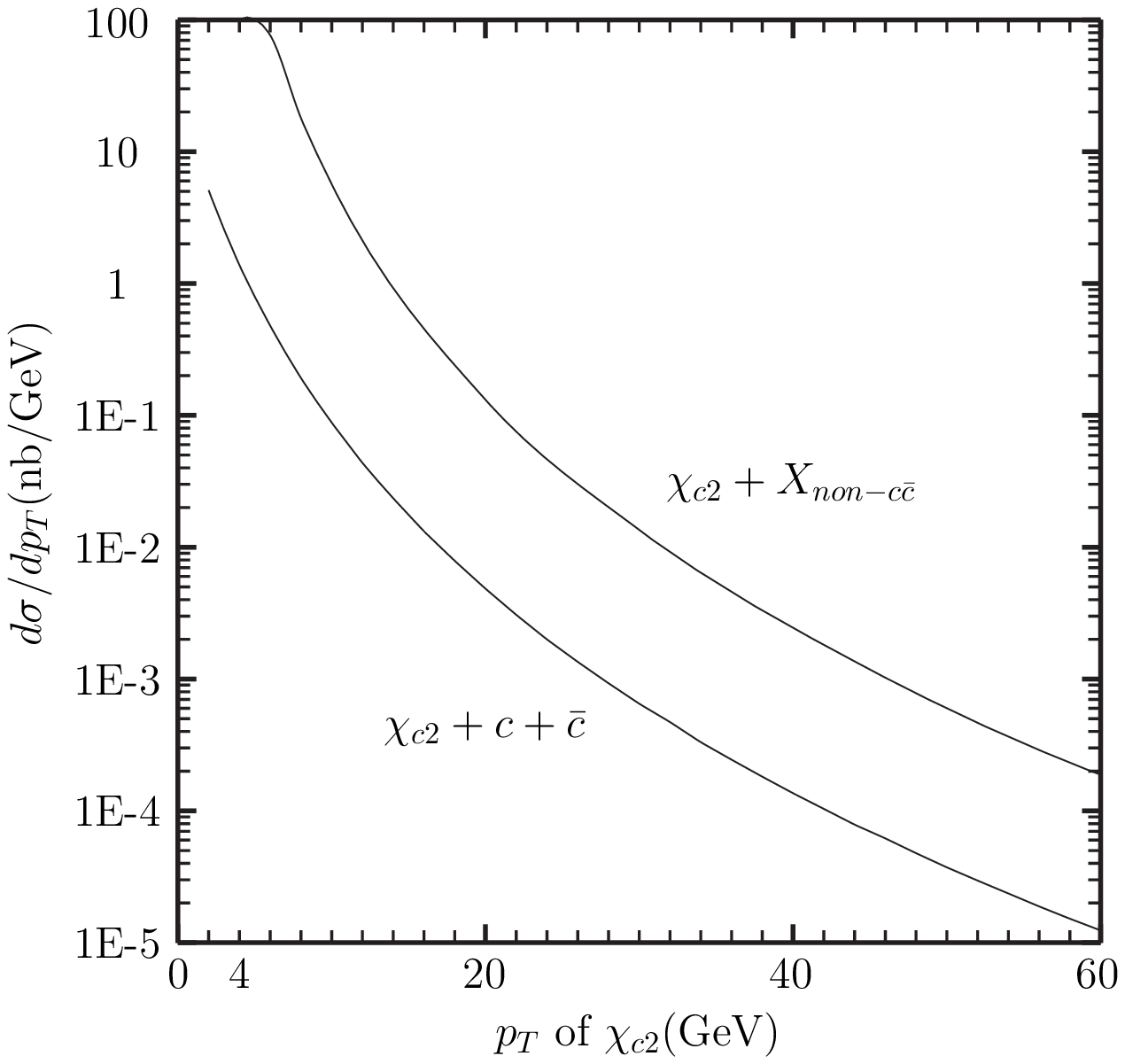}}
\end{center}
\caption{\label{fig:4}Comparison of the differential cross sections
of $\chi_{cJ}+c+\bar{c}$ with the NLO results of
$\chi_{cJ}+X_{non-c\bar{c}}$ at the Tevatron with
$\sqrt{s}=1.96\text{GeV}$ and rapidity cut $|y_{\chi_c}|<1$. }
\end{figure}

In Fig. \ref{fig:4}, we depict the differential cross section for
$\chi_c+c+\bar{c}$ and the NLO result for $\chi_c+X_ {non-c\bar{c}}$
\cite{Yan-Qing Ma:2010} as comparison. We find that the contribution
from $\chi_c+c+\bar{c}$ is about two orders of magnitude smaller
than $\chi_c+X_{non-c\bar{c}}$ at small $p_T$. The fraction of
$\chi_c+c+\bar{c}$ in total $\chi_c+X$ increases gradually and
reaches at most 20\% at $p_T$ as large as $60 \text{GeV}$.
The smallness of the $p\bar p\to\chi_c+c+\bar{c}$ fraction lies in
the fact that one of the main sources of $\chi_c+X_ {non-c\bar{c}}$
is the CO, which scales as $\frac{1}{p_T^4}$ and begins its
contribution at order $\alpha_s^3$, while the dominant contribution
of $\chi_c+c+\bar{c}$ is suppressed by both $\alpha_s$ and phase
space. Based on this analysis, we may conclude that
$\chi_c+c+\bar{c}$ has negligible influence on the $\chi_c$
inclusive production. As a result, fitting the ratio $R_{\chi_c}$
measured by CDF\cite{Abulencia:2007PRL} can not be improved by
including $p\bar p\to\chi_c+c+\bar{c}$ as compared to the $p\bar
p\to\chi_c+X_ {non-c\bar{c}}$ result\cite{Yan-Qing Ma:2010}.

\begin{figure}
\includegraphics[width=7.5cm]{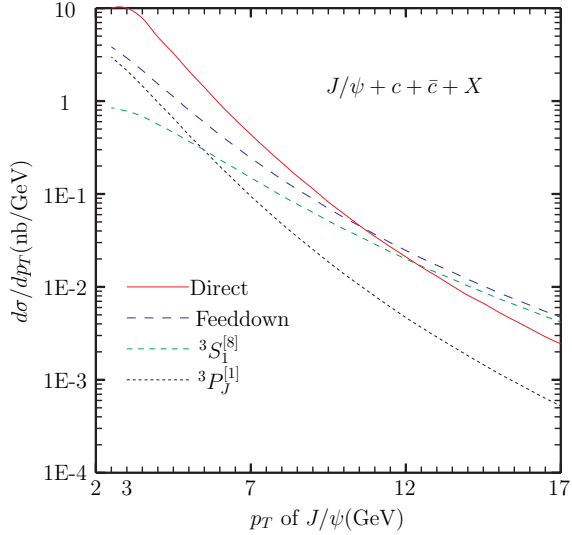}
\caption{\label{fig:5}Contribution of $\chi_{cJ}$ feeddown to prompt
$J/\psi+c+\bar{c}+X$ production at the Tevatron with rapidity cut
$|y_{\chi_c}|<0.6$. The dotted line denotes the contribution from CS
$\chi_{cJ}$ feeddown, the short dashed line denotes that from CO
$\chi_{cJ}$ feeddown, the long dashed line denotes that from CO+CS
$\chi_{cJ}$ feeddown, and the solid line is the direct
$J/\psi+c+\bar{c}$ contribution(from \cite{Artoisenet:2007PLB}).}
\end{figure}

In the NLO prediction\cite{Yan-Qing Ma:2010}, the feeddown
contribution of $\chi_{cJ}+X_{non-c \bar{c}}$ possesses about 30\%
of the prompt $J/\psi$ production rates at $p_T=20\text{GeV}$ at the
Tevatron, and it can give a great influence on $J/\psi$ prompt
production. Thus we also evaluate the feeddown contribution of
$\chi_{cJ}$ to $J/\psi+c+\bar{c}$  to see whether this contribution
is also large. In the calculation, we ignore the difference between
$p_T$ of $J/\psi$ and $\chi_c$. Note that the feeddown from
$\chi_{cJ}$ may have important influence on prompt $J/\psi$'s
polarization. This effect relies on $\chi_{cJ}$'s polarized
production rates and also the helicity amplitudes of $\chi_c$
radiative decays to polarized $J/\psi$. The related formula can be
found in Ref.\cite{Kramer:2003PRD}.
In this work, we only consider the unpolarized $\chi_c+c\bar{c}$
production but ignore the polarization effects.  The branching
ratios for $\chi_{cJ}$ radiative decays to $J/\psi$ are
$Br(\chi_{c0}\rightarrow J/\psi+\gamma)\approx 0.013$,
$Br(\chi_{c1}\rightarrow J/\psi+\gamma)\approx 0.36$,
$Br(\chi_{c2}\rightarrow J/\psi+\gamma)\approx 0.20$
respectively\cite{Yao:2006}. In Fig. \ref{fig:5},
we give the feeddown contribution from $\chi_{cJ}$  as a function of
$p_T$. By comparing it with the direct production, we can see that
the feeddown contribution from $\chi_{cJ}$ is small in the low $p_T$
region, but it is about a factor of 2 greater than direct
$J/\psi+c+\bar{c}$ contribution when $p_T>15\text{GeV}$. The turning
point is at $p_T\approx 9 \text{GeV}$ where the feeddown
contribution begins to exceed the direct contribution. For
$\chi_c+c+\bar{c}$ is dominated by the ${}^3S_1^{[8]}$ channel at
large $p_T$, one may anticipate that the CO $J/\psi+c+\bar{c}$
contribution from ${}^3S_1^{[8]}$ may also play an important role in
the direct $J/\psi+c+\bar{c}$ production. However, the magnitude
depends on the size of $\langle O^{J/\psi}({}^3S_1^{[8]})\rangle $.
In a recent work\cite{Ma:2010}, the authors find $\langle
O^{J/\psi}({}^3S_1^{[8]})\rangle $ might be small. As a result,
$\chi_c$ feeddown could become the main source for prompt production
of $J/\psi+c+\bar{c}$, if $\langle O^{J/\psi}({}^3S_1^{[8]})\rangle
$ is small. Thus when measuring the production cross sections for
prompt $J/\psi+c+\bar{c}$ at hadron colliders, the feeddown effect
from $\chi_{c}+c+\bar{c}$ can be very important and should be taken
into consideration.   We also note that this situation is different
from that at B factories, where the $\chi_{cJ}+c+\bar{c}$ production
rates in $e^+e^-$ annihilation for both CS and CO  are very small,
and their feeddown contributions to $J/\psi$ are negligible,
therefore, the NLO $J/\psi+c+\bar{c}$ theoretical results (including
direct and $\psi(2S)$ feeddown contributions) are basically
consistent with experimental production rates. At hadron colliders,
however, the $\chi_{cJ}+c+\bar{c}$ feeddown effect becomes more
important.

\begin{figure}[!]
\begin{center}
\scalebox{0.55}{\includegraphics{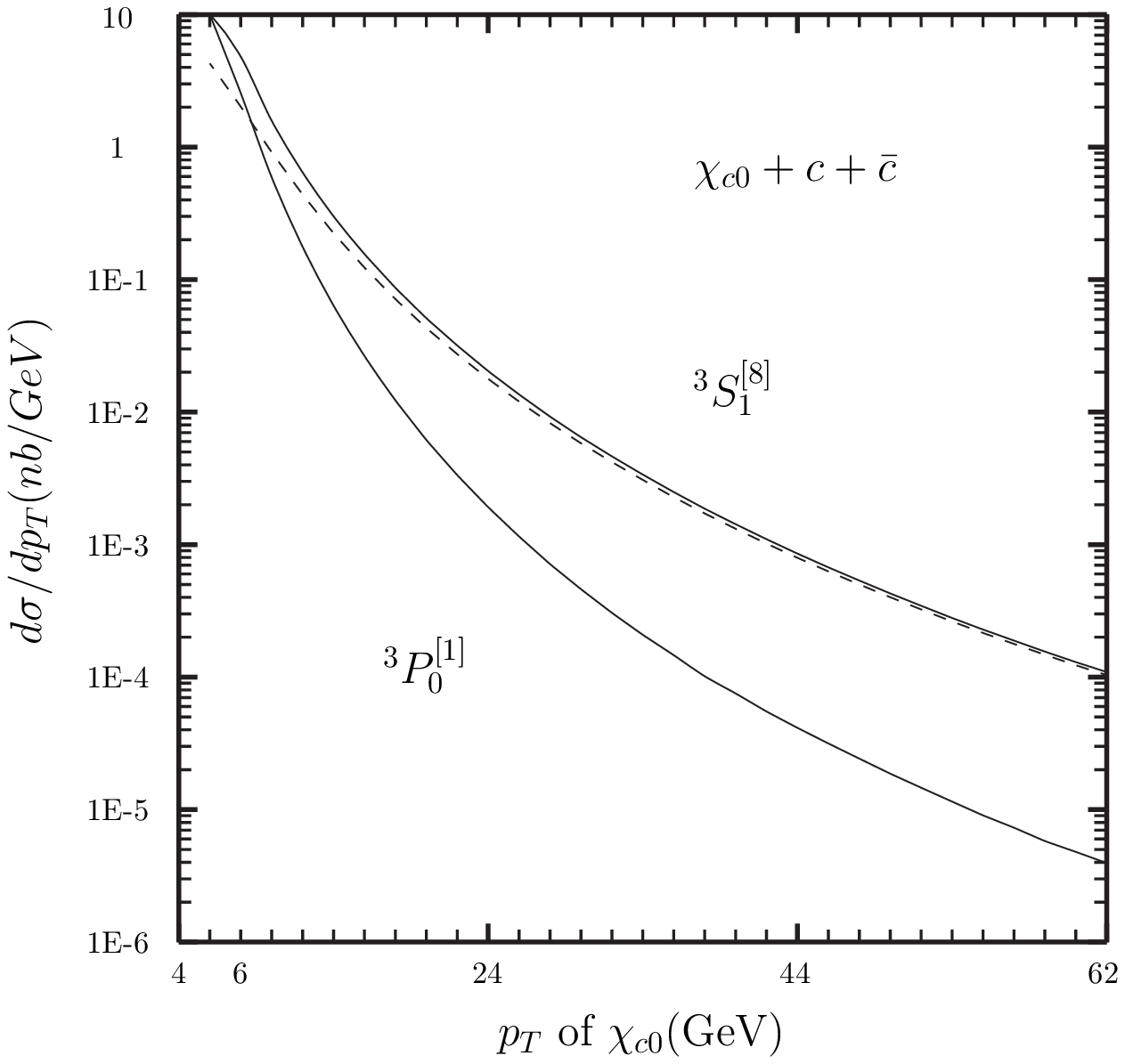}}\\
\scalebox{0.55}{\includegraphics{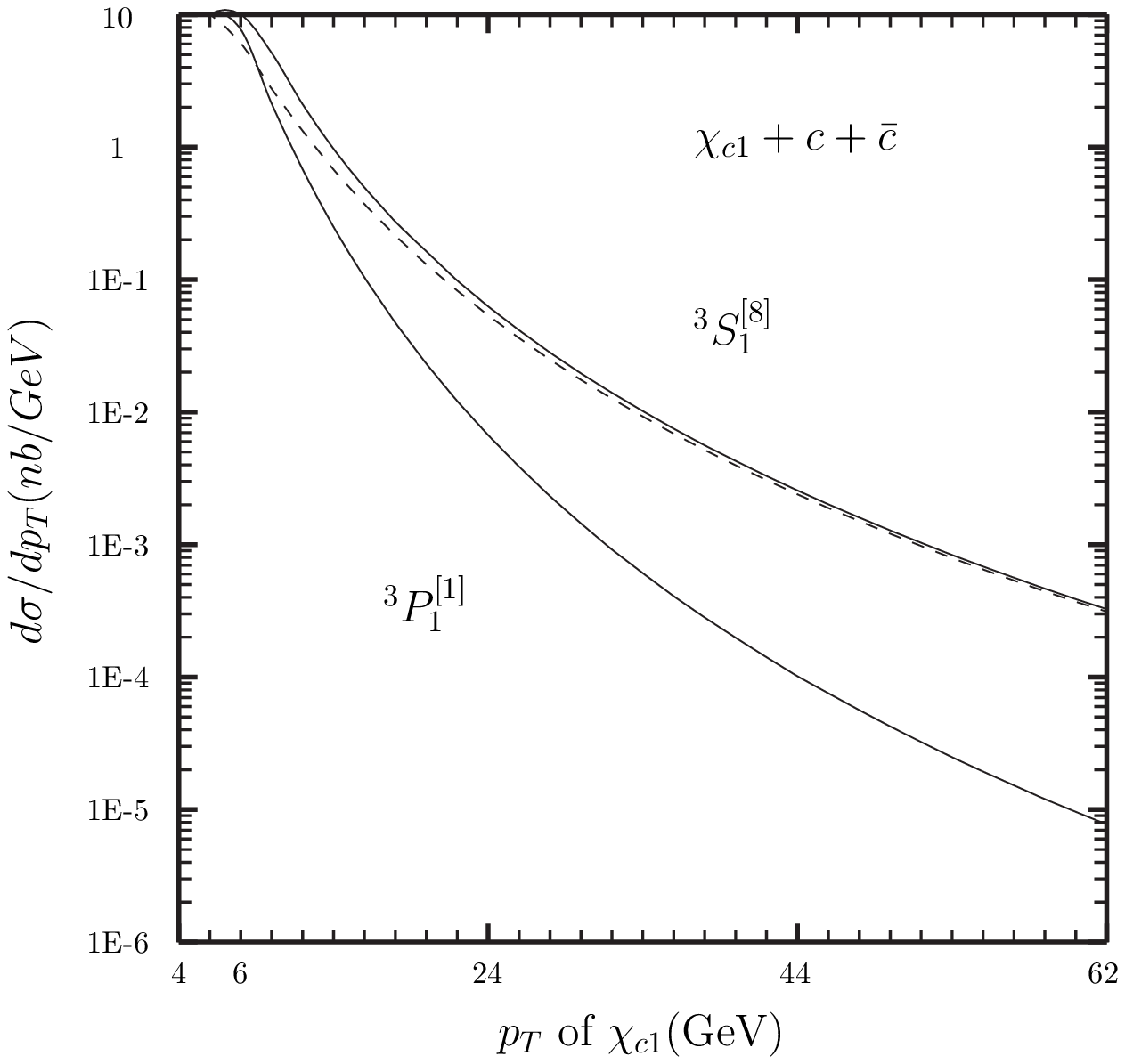}}\\
\scalebox{0.55}{\includegraphics{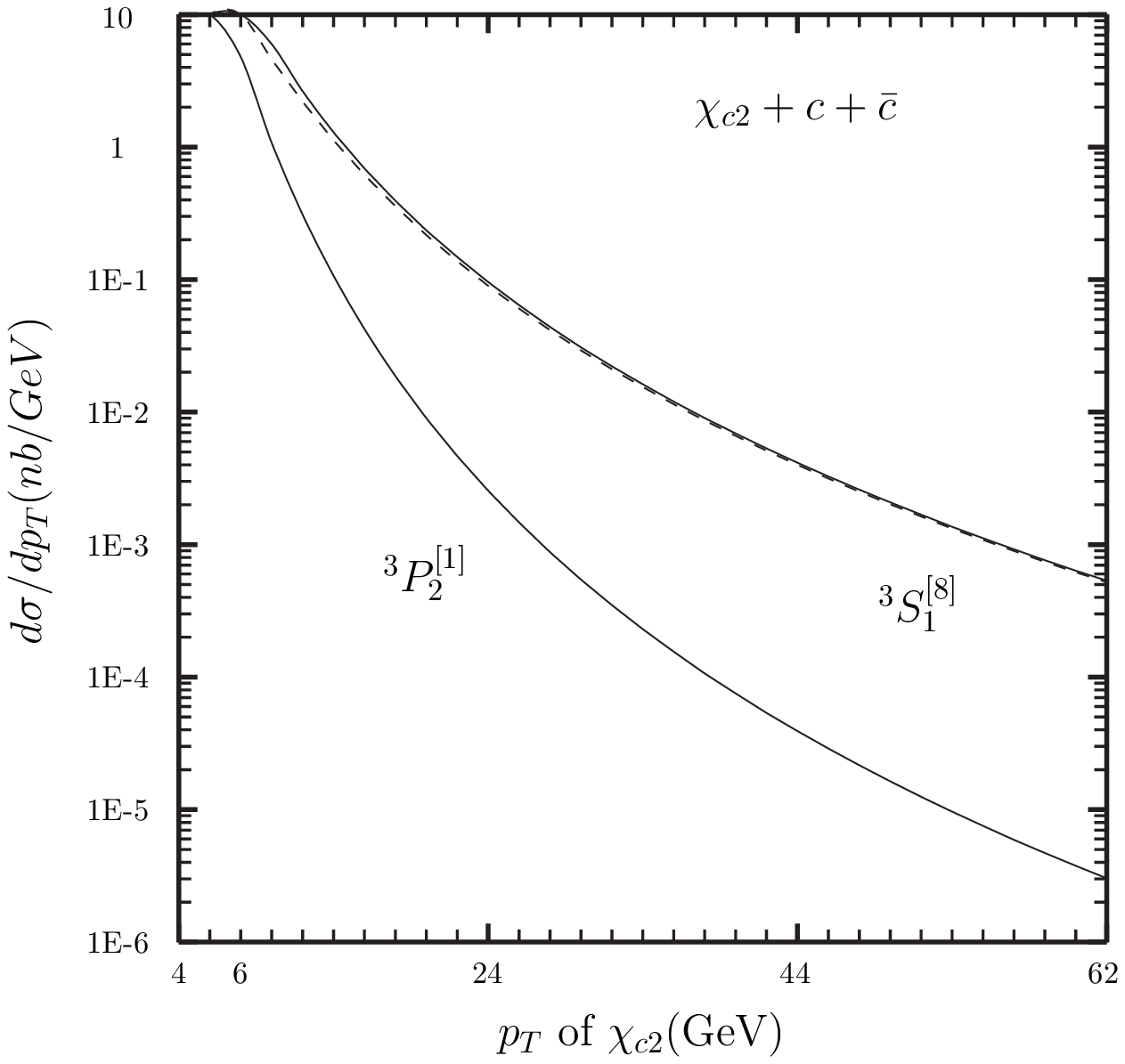}}
\end{center}
\caption{\label{fig:6}Differential cross sections of
$\chi_{cJ}+c+\bar{c}$ as functions of transverse momentum at the LHC
with $\sqrt{s}=7\text{TeV}$ and rapidity cut $|y_{\chi_{cJ}}|<2.4$.
The dashed line denotes CO contribution, and the solid line denotes
CS contribution. }
\end{figure}
\begin{figure}[!]
\begin{center}
\scalebox{0.55}{\includegraphics{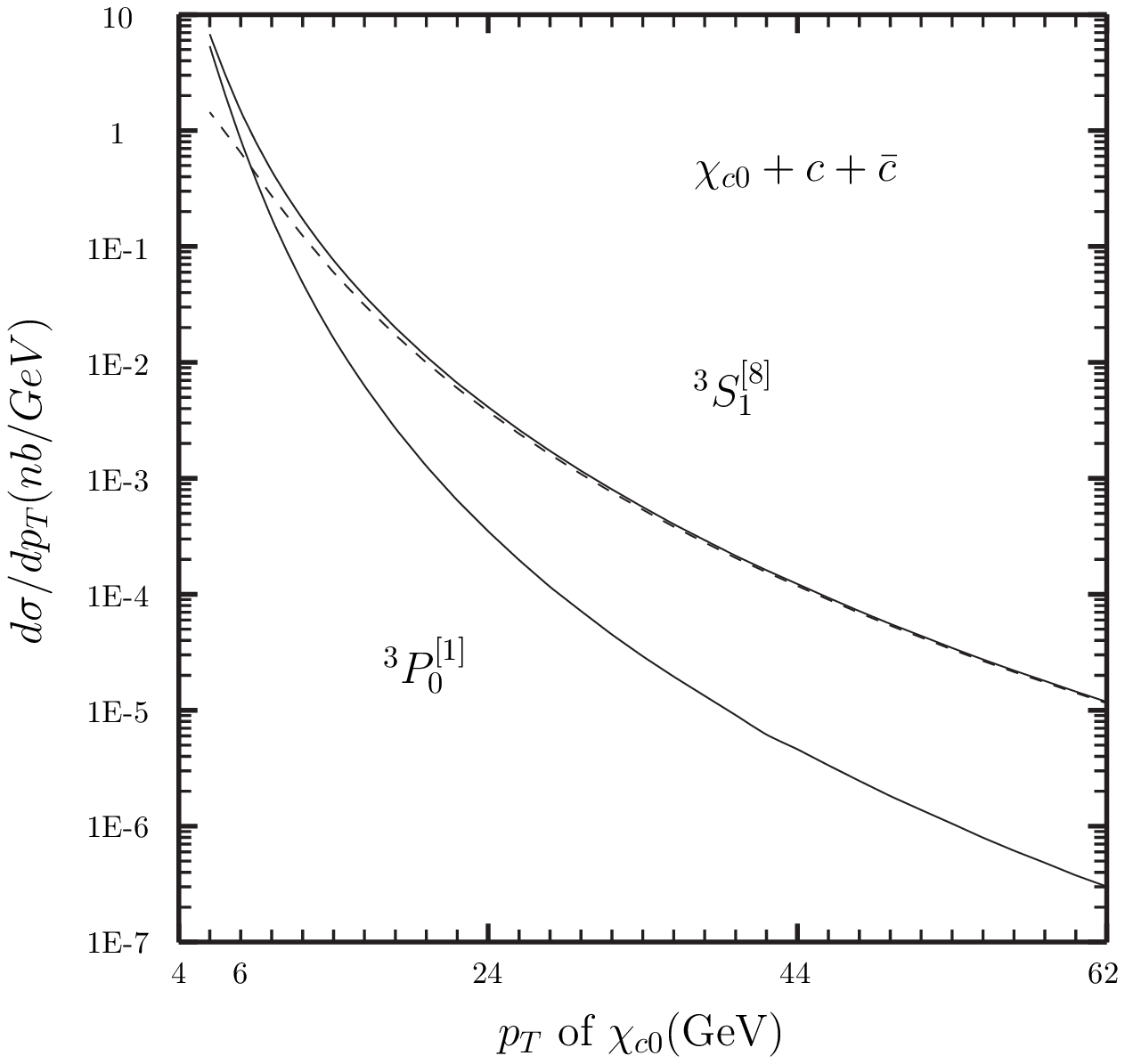}}\\
\scalebox{0.55}{\includegraphics{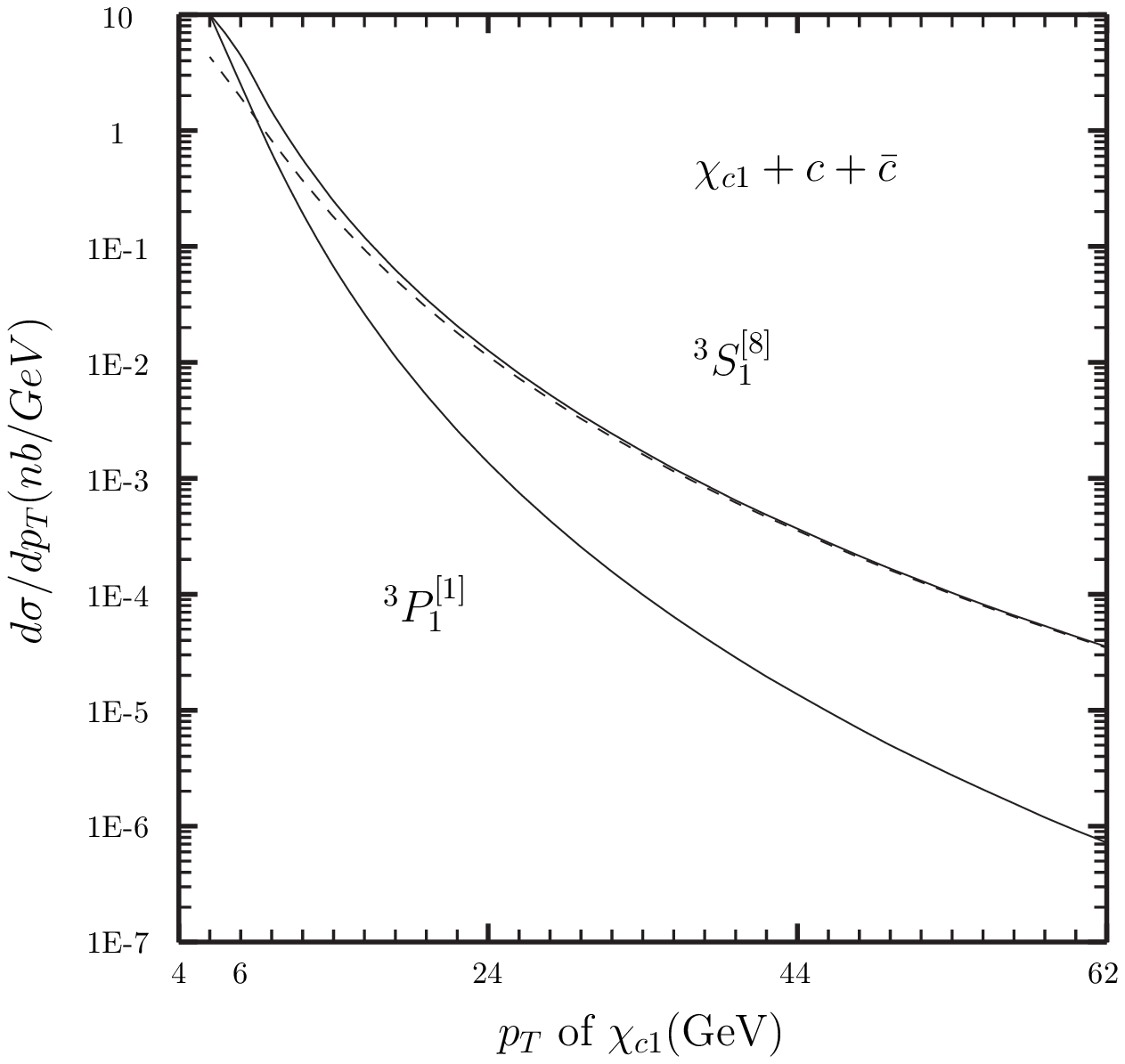}}\\
\scalebox{0.55}{\includegraphics{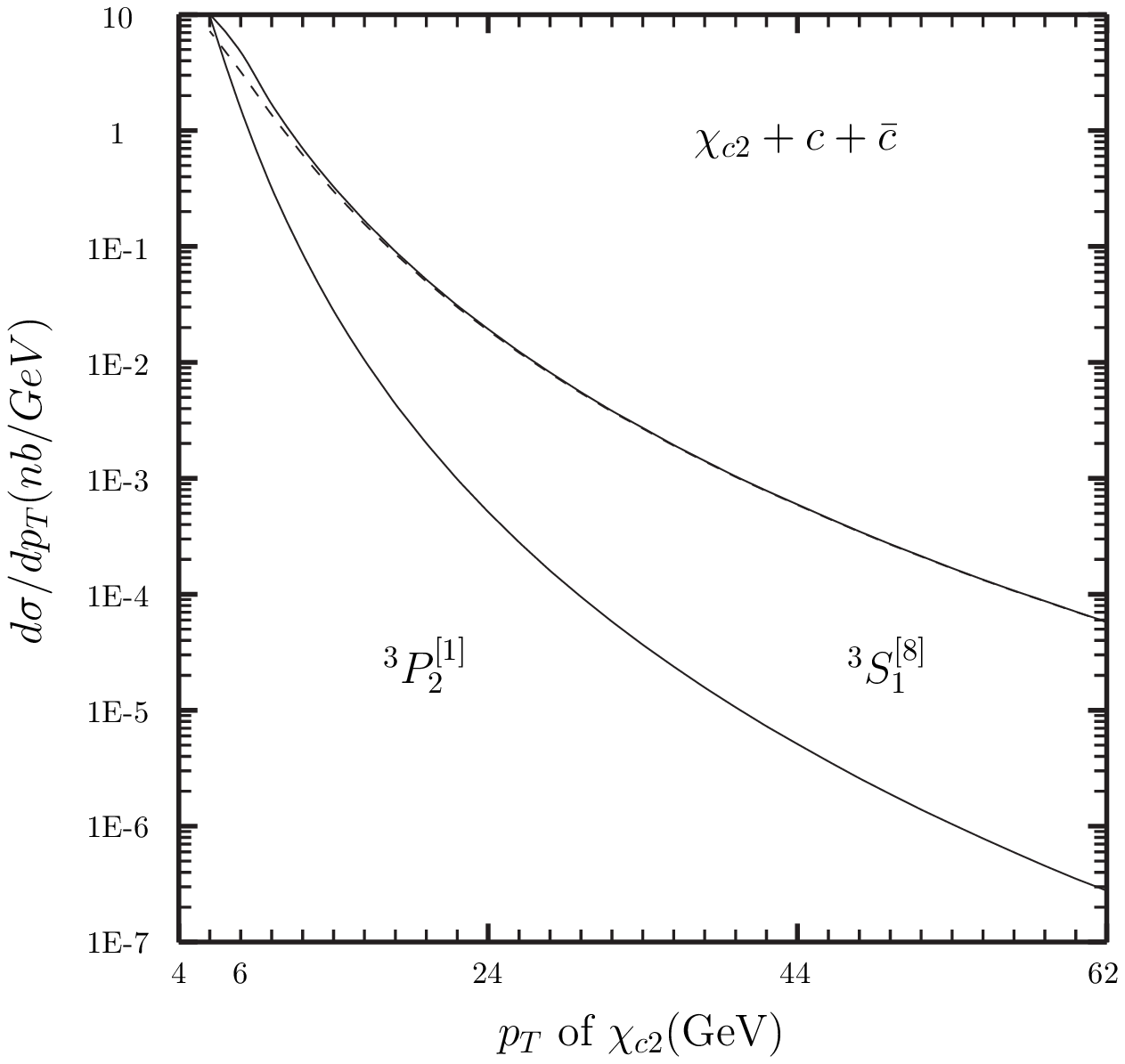}}
\end{center}
\caption{\label{fig:7}Differential cross sections of
$\chi_{cJ}+c+\bar{c}$ as functions of transverse momentum at the LHC
with $\sqrt{s}=7\text{TeV}$ and rapidity cut $2<y_{\chi_{cJ}}<4.5$.
The dashed line denotes CO contribution, and the solid line denotes
CS contribution. }
\end{figure}
\begin{figure}[!]
\begin{center}
\scalebox{0.55}{\includegraphics{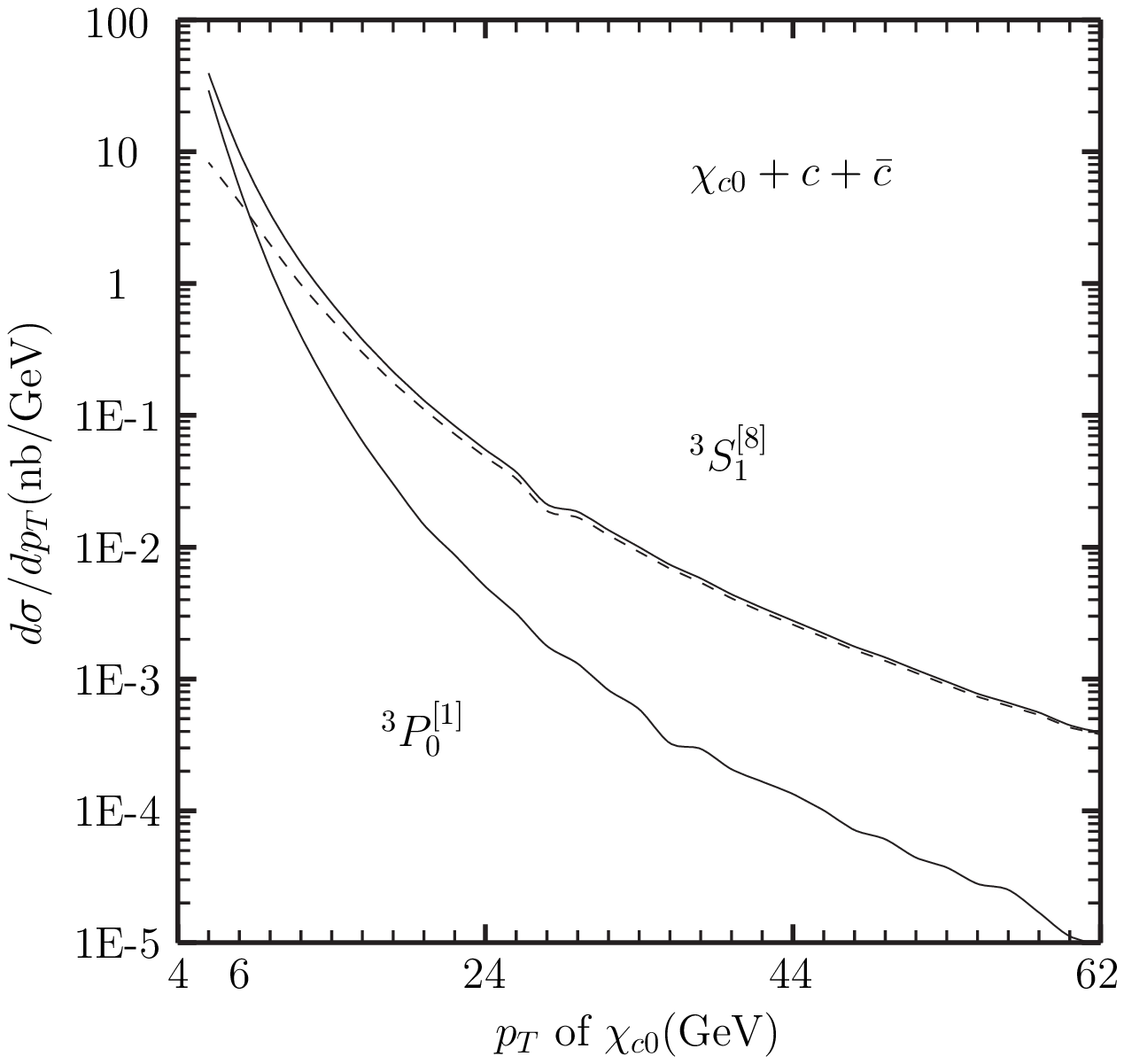}}\\
\scalebox{0.55}{\includegraphics{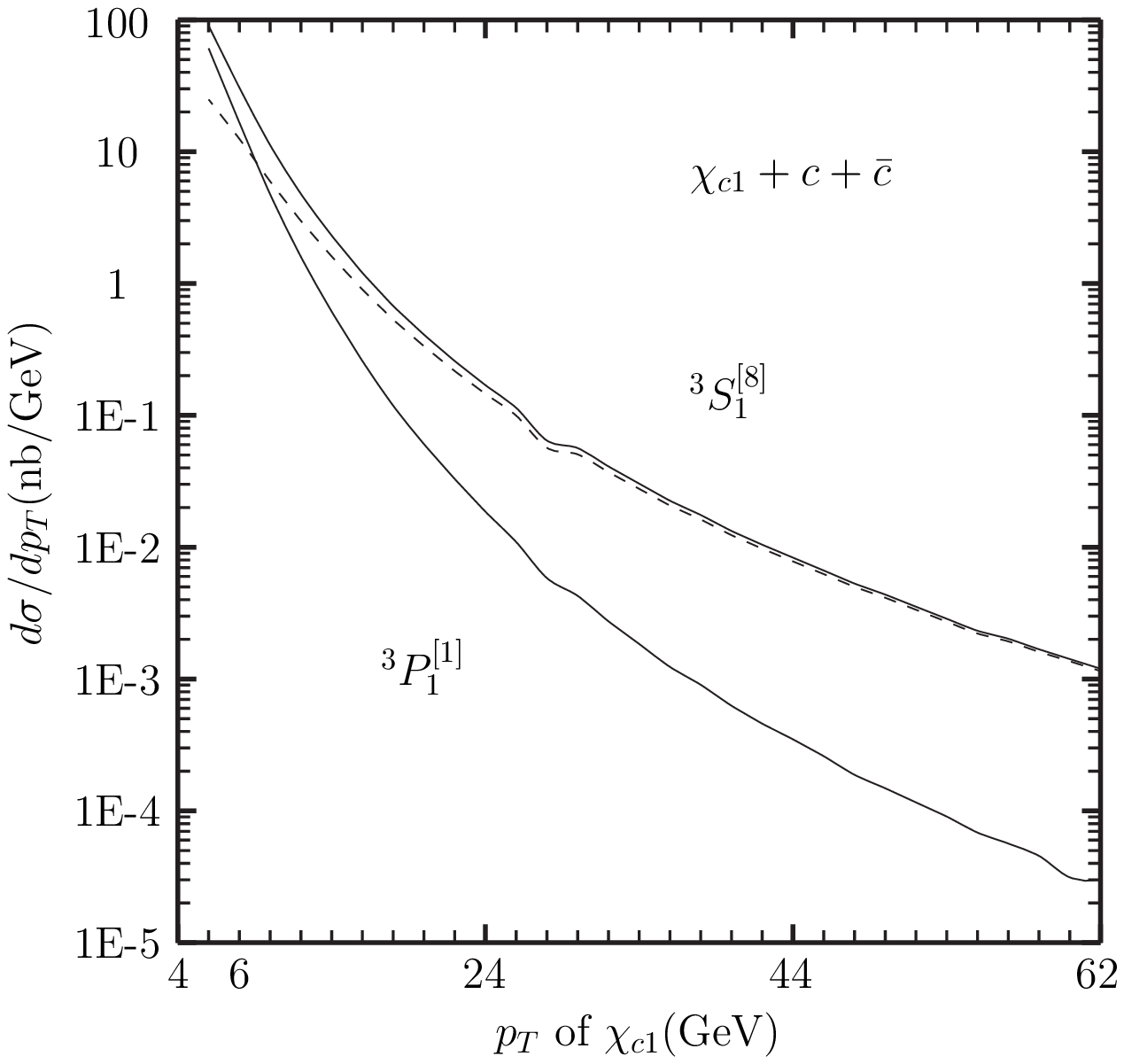}}\\
\scalebox{0.55}{\includegraphics{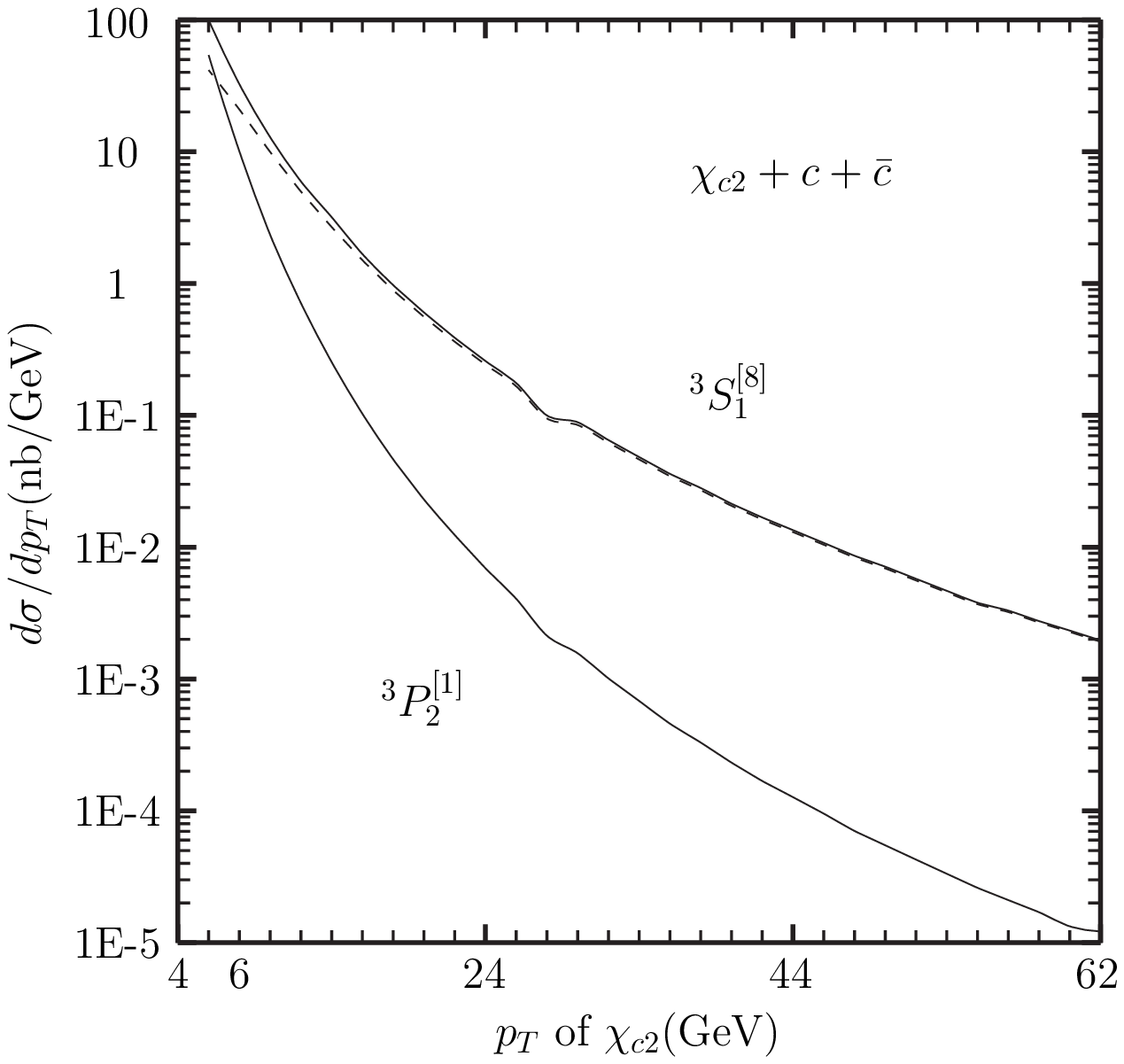}}
\end{center}
\caption{\label{fig:8}Differential cross sections of
$\chi_{cJ}+c+\bar{c}$ as functions of transverse momentum at the LHC
with $\sqrt{s}=14\text{TeV}$ and rapidity cut $|y_{\chi_{cJ}}|<3.0$.
The dashed line denotes CO contribution, and the solid line denotes
CS contribution. }
\end{figure}

We also give the prediction of $\chi_{cJ}+c+\bar{c}$ associated
production at the LHC with $\sqrt{s}=7\text{TeV}$. For the CMS
detector, the rapidity cut is $|y_{\chi_cJ}|<2.4$ and for the LHCb
detector, the rapidity cut is chosen as $2<y_{\chi_cJ}<4.5$. The
results are shown in Fig. \ref{fig:6} and Fig. \ref{fig:7} including
the CO contribution, CS contribution and the total differential
cross sections. We find that the differential cross sections for
$\chi_{cJ}+c+\bar{c}$ associated production at the LHC show a
similar behavior as that at the Tevatron: the CO contribution is
much larger than CS contribution in almost all $p_T$ region. So
$\chi_{cJ}$ associated production cross section is dominated by the
CO contribution. As a result, measuring $\chi_{cJ}+c+\bar{c}$
production can be used to determine the CO matrix element $\langle
O^{\chi_{c0}}[{}^3S_1^{[8]}]\rangle$. Predictions for LHC with
$\sqrt{s}=14\text{TeV}$ and $|y_{\chi_cJ}|<3$ are shown in Fig.
\ref{fig:8}.

\section*{\large{Summary}}
In this paper, we investigate the $\chi_{cJ}$ associated production
with a charm quark pair $p \bar{p}\rightarrow \chi_{cJ}+c+\bar{c}$
at hadron colliders in the framework of NRQCD.
By comparing the differential cross sections i.e. the transverse
momentum distributions for $\chi_{cJ}$ associated production, we
find that CO dominates the production rate at large $p_T$. Also, the
differential cross sections for associated $\chi_{cJ}$ production
are at least one order of magnitude smaller than the NLO result for
$\chi_{cJ}+X_{non-c\bar{c}}$. As a result, $\chi_{cJ}+c+\bar{c}$
production has negligible influence on the $R_{\chi_c}$ value
measured by the CDF collaboration. We also evaluate the $\chi_{cJ}$
feeddown contribution to prompt $J/\psi+c+\bar{c}$ production, and
find that the feeddown contribution is very large compared to direct
$J/\psi+c+\bar{c}$ production\cite{Artoisenet:2007PLB} at large
$p_T$, which illustrates the importance of $\chi_c$ feeddown effect
in the measurement for $J/\psi$ associated production cross sections
and polarization parameters. The fragmentation approximation is
analyzed and our conclusion is that the fragmentation contribution
is dominant for the CO channel, while for the CS channel the
fragmentation diagrams' contribution dominates over the total
differential cross section only at the $p_T \gtrsim 100\text{GeV}$
region.

Finally, we note that in the $\chi_{cJ}$ associated production, the
LO result in $\alpha_s$ has already contained the $1/{p_T}^4$ term,
which is the leading term in $1/{p_T}$ expansion at large ${p_T}$,
and high order corrections in $\alpha_s$ can at most give the
$1/{p_T}^4$ term, but suffer from suppressions due to extra powers
of $\alpha_s$. So we expect that high order corrections in
$\alpha_s$ can not significantly change the results obtained in this
work. Another notable result in this work is that the CO
contribution dominates over $\chi_{cJ}+c+\bar{c}$ production at
large $p_T$ (say, $p_T>7\text{GeV}$), therefore measuring the
process at hadron colliders, especially at the LHC, may provide
important information for the production mechanism of heavy
quarkonium, while the experiment itself may also be very interesting
and challenging in view of the complexity of the measurement.

\section*{\large{Acknowledgement}}
We thank Rong Li for valuable help in Fortran program of phase space
integration and Yu-Jie Zhang and Kai Wang and Ce Meng for helpful
discussions. This work was supported by the National Natural Science
Foundation of China (No.11021092, No.11075002) and the Ministry of
Science and Technology of China (No.2009CB825200).

\section*{\large{Appendix}}
We list all the fermion chains encountered in our calculation as
follows.
\begin{align}
f_0(q_1,q_2,\lambda_1,\lambda_2)=&\langle
q_{0\lambda_1}|(\slashed{q}_1+m)\gamma_{\delta}(\slashed{q}_2-m)|q_{0\lambda_2}\rangle
,\\
f_1(q_1,q_2,k,\lambda_1,\lambda_2,\lambda_3)=&\langle
q_{0\lambda_1}|(\slashed{q}_1+m)\gamma_{\delta}(\slashed{k}-\slashed{q}_2+m)\slashed{\epsilon}^{\lambda_3}(k,q_0)(\slashed{q}_2-m)|q_{0\lambda_2}\rangle
,\\
f_2(q_1,q_2,k,\lambda_1,\lambda_2,\lambda_3)=&\langle
q_{0\lambda_1}|(\slashed{q}_1+m)\slashed{\epsilon}^{\lambda_3}(k,q_0)(\slashed{q}_1-\slashed{k}+m)\gamma_{\delta}(\slashed{q}_2-m)|q_{0\lambda_2}\rangle
,\\
f_3(q_1,q_2,k,\lambda_1,\lambda_2,\lambda_3)=&\langle
q_{0\lambda_1}|(\slashed{q}_1+m)\slashed{\epsilon}^{\lambda_3}(\slashed{q}_2-m)|q_{0\lambda_2}\rangle
,\\
f_4(q_1,q_2,k_1,k_2,\lambda_1,\lambda_2,\lambda_3,\lambda_4)=&\langle q_{0\lambda_1}|(\slashed{q}_1+m)\gamma_{\delta}(\slashed{k}_1+\slashed{k}_2-\slashed{q}_2+m)\slashed{\epsilon}^{\lambda_3}(k_1,q_0)\nonumber\\
&(\slashed{k}_2-\slashed{q}_2+m)\slashed{\epsilon}^{\lambda_4}(k_2,q_0)(\slashed{q}_2-m)|q_{0\lambda_2}\rangle
,\\
f_5(q_1,q_2,k_1,k_2,\lambda_1,\lambda_2,\lambda_3,\lambda_4)=&\langle q_{0\lambda_1}|(\slashed{q}_1+m)\slashed{\epsilon}^{\lambda_3}(k_1,q_0)(\slashed{q}_1-\slashed{k}_1+m)\gamma_{\delta}\nonumber\\
&(\slashed{k}_2-\slashed{q}_2+m)\slashed{\epsilon}^{\lambda_4}(k_2,q_0)(\slashed{q}_2-m)|q_{0\lambda_2}\rangle
,\\
f_6(q_1,q_2,k_1,k_2,\lambda_1,\lambda_2,\lambda_3,\lambda_4)=&\langle q_{0\lambda_1}|(\slashed{q}_1+m)\slashed{\epsilon}^{\lambda_3}(k_1,q_0)(\slashed{q}_1-\slashed{k}_1+m)\slashed{\epsilon}^{\lambda_4}(k_2,q_0)\nonumber\\
&(\slashed{q}_1-\slashed{k}_1-\slashed{k}_2-m)\gamma_{\delta}(\slashed{q}_2-m)|q_{0\lambda_2}\rangle
,\\
f_7(q_1,q_2,k_1,k_2,\lambda_1,\lambda_2,\lambda_3,\lambda_4)=&\langle
q_{0\lambda_1}|(\slashed{q}_1+m)\slashed{\epsilon}^{\lambda_3}(k_1,q_0)\slashed{\epsilon}^{\lambda_4}(k_2,q_0)\gamma_{\delta}(\slashed{q}_2-m)|q_{0\lambda_2}\rangle
,\\
f_8(q_1,q_2,k_1,k_2,\lambda_1,\lambda_2,\lambda_3,\lambda_4)=&\langle
q_{0\lambda_1}|(\slashed{q}_1+m)\gamma_{\delta}\slashed{\epsilon}^{\lambda_3}(k_1,q_0)\slashed{\epsilon}^{\lambda_4}(k_2,q_0)(\slashed{q}_2-m)|q_{0\lambda_2}\rangle
,\\
f_9(q_1,q_2,k_1,k_2,\lambda_1,\lambda_2,\lambda_3,\lambda_4)=&\langle
q_{0\lambda_1}|(\slashed{q}_1+m)\slashed{\epsilon}^{\lambda_3}(k_1,q_0)(\slashed{q}_2-m)|q_{0\lambda_2}\rangle
,
 \end{align}
where $q_1^2=q_2^2=m_c^2$, $k$ ,$k_1$,$k_2$ and $q_0$ are light-like
vectors.

{}

\begin{thebibliography}{}
\bibitem{Abe:2002PRL}
K.Abe {\it et al.}[BELLE Collaboration], Phys. Rev. Lett.
\textbf{89}, 142001 (2002) [arXiv:hep-ex/0205104].

\bibitem{Pakhlov:2002PRD}
P. Pakhlov {\it et al.}[BELLE Collaboration], Phys. Rev.
\textbf{D79}, 071101 (2009) [arXiv: 0901.2775[hep-ex]].

\bibitem{Yu-Jie Zhang:2007PRL}
Yu-Jie Zhang, Kuang-Ta Chao, Phys. Rev. Lett. \textbf{98}, 092003
(2007) [arXiv: hep-ph/0611086]; Bin Gong, Jian-Xiong Wang, Phys.
Rev. \textbf{D80}, 054015 (2009) [arXiv: 0904.1103[hep-ph]].

\bibitem{Yan-Qing Ma:2009PRL}
Yan-Qing Ma, Yu-Jie Zhang and Kuang-Ta Chao, Phys. Rev.
Lett.\textbf{102}, 162002 (2009)[arXiv: 0812.5106[hep-ph]]; Bin
Gong, Jian-Xiong Wang, Phys. Rev. Lett. \textbf{102}, 162003 (2009)
[arXiv: 0901.0117[hep-ph]].

\bibitem{Yu-Jie Zhang:2009}
Yu-Jie Zhang, Yan-Qing Ma, Kai Wang and Kuang-Ta Chao, Phys. Rev.
\textbf{D81}, 034015 (2010) [arXiv: 0911.2166[hep-ph]].

\bibitem{Rong Li:2010}
  Rong Li and Jian-Xiong Wang,
  Phys.\ Rev.\  D {\bf 82}, 054006 (2010)
  [arXiv:1007.2368 [hep-ph]].

\bibitem{Abreu:1994PLB}
P. Abreu {\it et al.} [DELPHI Collaboration], Phys. Lett.
B\textbf{341}, 109 (1994); M. Wadhwa {\it et al.} [L3
Collaboration], Nucl. Phys. Proc. Suppl. \textbf{64}, 441 (1998); G.
Alexander {\it et al.} [OPAL Collaboration], Phys. Lett.
B\textbf{384}, 343 (1996); M. Acciarri {\it et al.} [L3
Collaboration], Phys. Lett. B\textbf{453}, 94 (1999).

\bibitem{Rong Li:2009PRD}
Rong Li and Kuang-Ta Chao, Phys.\ Rev.\  D {\bf 79}, 114020 (2009)
[arXiv: 0904.1643[hep-ph]].




\bibitem{Abdallah:2003PLB}
J. Abdallah {\it et al.} [DELPHI Collaboration], Phys. Lett.
B\textbf{565}, 76 (2003) [arXiv: hep-ex/0307049].

\bibitem{Artoisenet:2007PLB}
P. Artoisenet, J.P. Lansberg and F. Maltoni, Phys. Lett.
B\textbf{653}, 60-66 (2007)[arXiv:hep-ph/0703129].

\bibitem{zhiguoHe:2009PRD}
Zhi-Guo He, Rong Li and Jian-Xiong Wang, Phys. Rev. \textbf{D79},
094003 (2009) [arXiv: 0904.2069[hep-ph]].


\bibitem{Campbell:2007}
  J.~M.~Campbell, F.~Maltoni and F.~Tramontano,
  Phys.\ Rev.\ Lett.\  {\bf 98}, 252002 (2007)
  [arXiv:hep-ph/0703113]; Bin Gong, Jian-Xiong Wang, Phys. Rev. Lett.
\textbf{100}, 232001 (2008) [arXiv: 0802.3727[hep-ph]]; Bin Gong,
Jian-Xiong Wang, Phys. Rev. \textbf{D78}, 074011 (2008) [arXiv:
0805.2469[hep-ph]]; Rong Li, Jian-Xiong Wang, Phys. Lett.
B\textbf{672}, 51-55 (2009) [arXiv: 0811.0963[hep-ph]].


\bibitem{Abe: 1997PRL}
F. Abe {\it et al.} [CDF Collaboration], Phys. Rev. Lett.
\textbf{79}, 572, 578 (1997); A. A. Affolder {\it et al.} [CDF
Collaboration], Phys. Rev. Lett. \textbf{85}, 2886
(2000)[arXiv:hep-ex/0004027].

\bibitem{Zhi-Guo He:2009}
Zhi-Guo He, Jian-Xiong Wang, Phys. Rev. \textbf{D81}, 054030 (2010)
[arXiv:0911.0139[hep-ph]].

\bibitem{Bodwin:1994jh}
  G.~T.~Bodwin, E.~Braaten and G.~P.~Lepage,
  Phys.\ Rev.\  \textbf{D51}, 1125 (1995)
  [Erratum-ibid.\  \textbf{D55}, 5853 (1997)]
  [arXiv:hep-ph/9407339].

\bibitem{Abulencia:2007PRL}
A. Abulencia {\it et al.} [CDF Collaboration], Phys. Rev. Lett.
\textbf{98}, 232001 (2007) [arXiv: hep-ex/0703028].


\bibitem{Yan-Qing Ma:2010}
Yan-Qing Ma, Kai Wang and Kuang-Ta Chao,  Phys. Rev. D83, 111503(R)
(2011)[arXiv: 1002.3987[hep-ph]].

\bibitem{Hahn:2001}
T. Hahn, Comput. Phys. Commun. \textbf{140}, 418 (2001).

\bibitem{Chang:2004cpc}
Chao-Hsi Chang, Chafik Driouichi, Paula Eerola, Xing-Gang Wu,
Comput. Phys. Commun. \textbf{159}, 192-224 (2004)
[arXiv:hep-ph/0309120].

\bibitem{Maitre:2007}
D.Maitre and P.Mastrolia, Comput. Phys. Commun. \textbf{179},
501-574 (2008)[arXiv: 0710.5559[hep-ph]].

\bibitem{Dan:2009PRD}
Dan Li, Zhi-Guo He and Kuang-Ta Chao, Phys. Rev. \textbf{D80},
114014 (2009) [arXiv: 0910.4155[hep-ph]].

\bibitem{Chao-Hsi Chang:2004PRD}
Chao-Hsi Chang, Jian-Xiong Wang, Xing-Gang Wu, Phys.Rev.
\textbf{D70}, 114019 (2004) [arXiv:hep-ph/0409280].


\bibitem{Lepage:Vegas}
  G.~P.~Lepage,
  ``Vegas: An Adaptive Multidimensional Integration Program''.


\bibitem{Eichten:1995PRD}
E.J. Eichten and C. Quigg, Phys. Rev. \textbf{D52}, 1726 (1995).

\bibitem{Kramer:2003PRD}
B. A. Kniehl, G. Kramer and C. P. Palisoc, Phys. Rev. \textbf{D68},
114002 (2003) [arXiv: hep-ph/0307386]; B. A. Kniehl and J. Lee,
Phys. Rev. \textbf{D62}, 114027 (2000).

\bibitem{Yao:2006}
  K.~Nakamura {\it et al.}  [Particle Data Group],
  J.\ Phys.\ G {\bf 37}, 075021 (2010).

\bibitem{Ma:2010}
Yan-Qing Ma, Kai Wang, Kuang-Ta Chao, Phys. Rev. Lett. \textbf{106},
042002 (2011) [arXiv: 1009.3655[hep-ph]].



\end{thebibliography}
\end{document}